\documentclass[runningheads]{svmult}

\usepackage{makeidx}   
\usepackage{graphicx}  
\usepackage{subeqnar}  
\usepackage{multicol}  
\usepackage{physprbb}  
\makeindex             

\usepackage{amsfonts}   
\begin{document}
\title*{Physics of rotation in stellar models}
\toctitle{Physics of rotation in stellar models}
%
%
\titlerunning{Physics of rotation in stellar models}
%

\author{Georges Meynet}
\authorrunning{Georges Meynet}

\institute{Astronomical Observatory of Geneva University\\ 
CH-1290, Sauverny, Switzerland \\
E-mail: \texttt{georges.meynet@obs.unige.ch}}


\maketitle              

\begin{abstract}
\noindent In these lecture notes, we present the equations presently used in stellar interior models in order to
compute the effects of axial rotation. We discuss the hypotheses made. We
suggest that the effects of rotation might play a key role at low metallicity.
\end{abstract}

\section{Physics of rotation}

Axial rotation modifies the hydrostatic equilibrium configuration by adding a centrifugal acceleration term in the hydrostatic equation, induces many instabilities driving the transport of angular momentum and of chemical species in radiative zones and
changes the mass loss rates. In the present lecture notes we shall consider the case
of models without magnetic fields.

\subsection{Hydrostatic effects of rotation}

\subsubsection{The equations of stellar structure.}

In a rotating star, the equations of stellar structure
 need to be modified  \cite{KipTho70}.
The usual spherical coordinates must be replaced by new coordinates 
characterizing the equipotentials.
The classical method applies when  the effective
gravity can be derived from a potential
$\Psi = \Phi - \frac{1}{2} \Omega^2 r^2 \sin^2 \theta$,
 i.e. when the  problem is conservative. There, $\Phi$ is the
gravitational potential, $\Omega$ the angular velocity, $r$ the radius at the colatitude $\theta$.
If  the rotation law is shellular ({\it i.e.} such that $\Omega$ is constant on isobaric surfaces see below), the problem is non--conservative.
Most existing models of rotating stars apply, rather
inconsistently, the classical scheme by \cite{KipTho70}.
However, as shown by \cite{MM97},
the equations of stellar structure can still be 
written consistently, in term of a coordinate referring to 
the mass inside the isobaric surfaces\footnote{For shellular rotation, the shape of the isobaric surfaces are given by the same expression as the one giving the shape of the equipotentials in conservative cases provided some changes of variables are performed.} . Thus, the problem
of the stellar structure
of a differentially rotating star in a shellular rotation state
can be kept one--dimensional.

\subsubsection{The Roche model.}

In all the derivations, we shall use the Roche model, {\it i.e.} we approximate the gravitational potential by $GM_{\overline{r}}/\overline{r}$ where $M_{\overline{r}}$ is the mass inside the isobaric surface with a mean radius $\overline{r}$. The radius $\overline{r}$ which labels each isobaric surface is defined by $\overline{r}=\left( V_{\overline{r}}/(4/3 \pi)\right )^{1/3}$ where $V_{\overline{r}}$ is the volume (deformed by rotation) inside the isobaric surface considered.
Apart from the case
of extreme rotational velocities, the parameter $\overline{r}$ is close to  the average 
radius of an isobar, which is the radius at $P_2 (\cos \vartheta) =0$,
namely for $\vartheta = 54.7$ degrees. 

In the frame of the Roche model, the shape of a meridian at the surface of the star (which is
an isobaric surface) is given by couples of $R$ and $\theta$ values satisfying the following equation:
\begin{equation}
{GM \over R}+{1 \over 2}\Omega^2 R^2 \sin^2 \theta={GM \over R_{\rm p}},
\label{eq1}
\end{equation}
where $R$ is the radius at colatitude $\theta$, $\Omega$ the angular velocity, $M$ the mass inside the surface and $R_{\rm p}$, the polar radius. Thus the shape of the surface (as well as
of any isobaric surface inside the star) is determined by three parameters $M$, $\Omega$ and $R_{\rm p}$. The first two $M$ and $\Omega$ are independent variables. The third one is derived from the first two and the equations of stellar structure. Setting $x=\left({GM \over \Omega^2}  \right)^{-1/3} R$, one can write Eq.~\ref{eq1} (see \cite{KipTho70})
\begin{equation}
{1 \over x}+{1 \over 2}x^2 \sin^2 \theta={1 \over x_{\rm p}}.
\label{eq1b}
\end{equation}
With this change of variable, the shape of an equipotential is uniquely determined by only one parameter $x_{\rm p}$.

Setting
\begin{equation}
f={R_{\rm e} \over R_{\rm p}},
\label{eq2}
\end{equation}
where $R_{\rm e}$ is the equatorial radius, one easily obtains from Eq.~\ref{eq1} that
\begin{equation}
R_{\rm p}=\left({GM \over \Omega^2} \right)^{1/3} \left({2 (f-1)\over f^3}
\right)^{1/3}=\left({GM \over \Omega^2} \right)^{1/3} x_{\rm p}.
\label{eq3}
\end{equation}
The above equation relates the inverse of the oblateness $f$ to $R_{\rm p}$. 

\subsubsection{The von Zeipel theorem and its consequences.}

The von Zeipel theorem \cite{vZ24}  expresses that the radiative
 flux $\vec{F}$
at some colatitude $\vartheta$ in a rotating star is proportional 
to the local effective  gravity $\vec{g_\mathrm{eff}}$.
\cite{Ma99}
has generalized this theorem to the case of shellular rotation and 
the  expression of the flux $\vec{F}$ for a star with
angular velocity $\Omega$ on the isobaric stellar surface  is

\begin{equation}
\vec{F}  =  - \frac{L(P)}{4 \pi GM_{\star}}
\vec{g_{\rm{eff}}} [1 + \zeta(\vartheta)]  \quad {\mathrm{with}} \quad 
\label{Ma23}
\end{equation} 
 
\begin{equation}
M_{\star} = M \left( 1 - \frac{\Omega^2}
{2 \pi G \rho_{\rm{m}}}  \right)
\quad {\mathrm{and}} \quad
\label{Ma24}
\end{equation}

 \begin{equation}
\zeta(\vartheta) = \left[\left(1 - \frac{\chi_T}{\delta}\right) \Theta +
\frac{H_T}{\delta} \frac{d\Theta}{dr}\right] P_{2}(\cos \vartheta).
\label{Ma25}
\end{equation}

\noindent
There, $\rho_{\rm{m}}$ is the internal average density,
 $\chi = 4acT^3/(3 \kappa \rho)$ and $\chi_{T}$
is the partial derivative with respect to T.  The quantity 
$\Theta$ is defined by 
$\Theta = \frac{\tilde{\rho}}{\bar{\rho}}$, 
i.e. the ratio of the horizontal density fluctuation
to the average density on the isobar \cite{Z92}.
One has the thermodynamic coefficients $\delta = - (\partial \ln\rho /
 \partial \ln T)_{P, \mu}$, $H_{T}$ is the temperature scale height.
 The term $\zeta(\vartheta)$, 
which expresses the deviations of the von Zeipel theorem due to the 
baroclinicity of the star,
is generally very small, (cf. \cite{Ma99}).

Let us emphasize that the flux is proportional to $\vec{g_\mathrm{eff}}$
and not to $\vec{g_\mathrm{tot}}$. This results from the fact that
the equation of hydrostatic equilibrium is 
$\frac{\vec{\nabla} P}{\rho} = - \vec{g_\mathrm{eff}}$.
The effect
of radiation pressure is already counted in the 
expression of $P$, which is the total pressure.
We may call $M_{\star}$ the effective mass, i.e. the mass
reduced by  the centrifugal
force. This is the complete form of the von Zeipel theorem in a 
differentially rotating star with shellular rotation, whether or not 
one is close to the Eddington limit.

This theorem has numerous consequences. Some of them are discussed below.
\vskip 2mm
\noindent{\underline{The position of the star in the HR diagram}} 
\vskip 2mm
A fast rotating star has stronger radiative fluxes at the pole than at the equator. Therefore the position of such a star in the HR
diagram will depend of the angle between the line of sight and the rotational axis (inclination angle). If for instance that angle is 90 degrees, a great part of the light will come from the equatorial belt characterized by lower radiative fluxes and cooler effective temperatures, while when the star is observed pole-on
most of the light will come from the hot polar region characterized by stronger fluxes and higher effective temperatures. Thus the perceived luminosity and effective temperature (and also effective gravity) of a star depend on the inclination angle.
This has to be kept in mind when comparisons
are made with observed quantities. Computations of the effect of the inclination angle on 
the emergent luminosity, colors and spectrum have been performed
by \cite{MP70}. The effect of the inclination angle 
on the determination of the effective gravity is discussed in \cite{HG06}.
In general, a theoretical evolutionary track is given in term of total luminosity
and of an average effective temperature defined by $T_{\rm eff}^4= L/(\sigma S(\Omega))$, 
where $\sigma$ is Stefan's constant and $S(\Omega)$ the total actual
stellar surface. The total luminosity (corresponding to the integrated
flux over the surface) does not depend on the angle of view, but
cannot be directly compared to the ``observed'' luminosity deduced from the apparent luminosity
coming from the hemisphere directed toward us. Let us note however that for surface velocities inferior to about 70\% of the critical velocity 
these effects remain quite modest. 
As a numerical example,
the ratio ($T_{\rm eff}$(pole)-$T_{\rm eff}$(equator))/$T_{\rm eff}$(equator) becomes superior to 10\% only for $\omega > 0.7$. At break-up, the effective temperature of the polar region is about a factor two higher than that of the equatorial one.
\vskip 2mm
\noindent{\underline{The Eddington luminosity}}
\vskip 2mm
Let us express  the total gravity at some colatitude 
$\vartheta$, taking into account the
radiative acceleration (cf. \cite{Ma99})

\begin{equation}
 \vec{g_\mathrm{rad}} = \frac{1}{\rho} \vec{\nabla} P_\mathrm{rad} = \frac{\kappa(\vartheta)\vec{F}}{c} \; ,
\end{equation}

\noindent
thus one has

\begin{eqnarray}
\vec{g_\mathrm{tot}}    =\vec{g_\mathrm{eff}}+\vec{g_\mathrm{rad}}=
\vec{g_\mathrm{eff}}+\frac{\kappa(\vartheta)\vec{F}}{c}
\end{eqnarray}

\noindent
The rotation effects appear both in $\vec{g_\mathrm{eff}}$ and in $\vec{F}$.
We may also consider the local limiting flux.
The condition $\vec{g_\mathrm{tot}}= \vec{0}$
 allows us to
define a limiting flux,

\begin{equation}
\vec{F_{\mathrm{lim}}}(\vartheta) = - \frac{c}{\kappa(\vartheta)}  
\vec{g_{\mathrm{eff}}}(\vartheta) \; .
\end{equation}

\noindent
From that we may define the ratio  $\Gamma_{\Omega}(\vartheta)$
of the  actual flux (see Eq.~\ref{Ma23})
$F(\vartheta)$ to the limiting local flux in a rotating star,

\begin{eqnarray}
\Gamma_{\Omega}(\vartheta) =
\frac{\vec{F}(\vartheta)}{\vec{F_{\mathrm{lim}}}(\vartheta)}=
\frac{ \kappa (\vartheta) \; L(P)[1+\zeta (\vartheta)]}{4 \pi 
cGM \left( 1 - \frac{\Omega^2}{2 \pi G \rho_{\rm{m}}}  \right) }\; .
\label{Ma29}
\end{eqnarray}

\noindent
 As a matter of fact,
 $\Gamma_{\Omega}(\vartheta)$ is the local Eddington ratio and
\begin{eqnarray}
L_{\rm Edd} =
\frac{4 \pi 
cGM \left( 1 - \frac{\Omega^2}{2 \pi G \rho_{\rm{m}}}  \right) }{\kappa (\vartheta) \; [1+\zeta (\vartheta)] }
\label{Lmax}
\end{eqnarray}
is the local Eddington luminosity.
For a certain angular velocity $\Omega$ on the 
isobaric surface, the maximum permitted luminosity of a star is
reduced by rotation, with respect to the usual Eddington limit.
In the above relation, $\kappa(\vartheta)$ is the largest value of
the opacity on the surface of the rotating star.
For O--type stars with photospheric opacities dominated
by electron scattering, the opacity $\kappa$ is the same
everywhere on the star.

For zero rotation, the usual expressions are found: $\Gamma_{\Omega}(\vartheta) = \Gamma=\frac{\kappa L}{4 \pi c GM}$
and $L_{\rm Edd} =\frac{4 \pi 
cGM }{\kappa}$ .
\vskip 2mm
\noindent{\underline{Critical limits.}}
\vskip 2mm
Critical limits correspond to values of respectively the luminosity and/or the velocity
which impose that the total gravity become equal to zero at least at some peculiar places at the surface.
We may identify different limits \cite{mm6}:

\vspace{2mm}
\noindent
-- We speak of the Eddington or $\Gamma$--limit,
when rotation effects can be neglected  and 
$\vec{g_\mathrm{rad}} + \vec{g_\mathrm{grav}} = \vec{0}$\footnote{$\vec{g_\mathrm{grav}}$ is the gravitational acceleration.}, 
which implies that

\begin{equation}
\Gamma = \frac{\kappa L}{4 \pi c GM}  \; \rightarrow  \; 1.
\end{equation}

\noindent In that case 
$L = L_{\rm Edd}=4 \pi c GM/\kappa$.
The opacity $\kappa$ considered here is the total opacity.

\vspace{2mm}
\noindent
--The critical velocity or $\Omega$--limit is reached for a star
with an angular velocity $\Omega$ at the surface, when the effective
gravity  $\vec{g_\mathrm{eff}} = \vec{g_\mathrm{grav}} +
 \vec{g_\mathrm{rot}} = \vec{0}$ and in addition
when radiation pressure effects can be neglected.

\vspace{2mm}
\noindent
--\textit{The $\Omega \Gamma$--limit is  reached when the total gravity
$\vec{g_\mathrm{tot}} = \vec{0}$, with significant effects of both
rotation and radiation.} 
This is the  general case. It should
lead to the two above cases in their respective limits.

Using relation \ref{Ma29}, we may
write the expression for the total gravity  as

\begin{equation}
\vec{g_\mathrm{tot}} = \vec{g_\mathrm{eff}}
\left[ 1 - \Gamma_{\Omega}(\vartheta) \right] \; .
\end{equation}

\noindent
This  shows that the 
expression for the total acceleration in a rotating star is 
similar to the usual one, except  that $\Gamma$ is 
replaced by the local value  $\Gamma_{\Omega}(\vartheta)$. 
Indeed, contrarily to expressions such as $\vec{g_\mathrm{tot}} = \vec{g_\mathrm{eff}} \left( 1 - \Gamma \right)$
often found in literature, we see that the appropriate Eddington factor given by Eq.~\ref{Ma29} also depends on the angular velocity $\Omega$ on the 
isobaric surface. 

From Eq.~\ref{Ma29}, we note that over the surface of a rotating star, which
has a varying gravity and T$_\mathrm{eff}$,  $\Gamma_{\Omega}(\vartheta)$
is the highest at the latitude where $\kappa(\vartheta)$ is the largest,
(if we neglect the effects of $\zeta (\vartheta)$, which is justified 
in general). If the opacity increases with decreasing T as in  hot
stars, the opacity is the highest at the equator and there the limit
 $\Gamma_{\Omega}(\vartheta) = 1 $ may by reached first. Thus, it is
to be stressed that if the limit  $\Gamma_{\Omega}(\vartheta) = 1 $ 
happens to be met at the equator, it is not because 
$\vec{g_\mathrm{eff}}$ is the lowest there, but because the 
opacity is the highest ! The reason for no direct dependence
on $\vec{g_\mathrm{eff}}$ is because both terms $\vec{g_\mathrm{eff}}$
cancel each other in the expression~\ref{Ma29} of the flux ratio.

The critical velocity is reached when somewhere on the star one has
$\vec{g_\mathrm{tot}}=\vec{0}$, i.e. 

\begin{equation}
 \vec{g_\mathrm{eff}} \;\left[1 - \Gamma_{\Omega}(\vartheta)
 \right] = \vec{0} .
 \label{Ma313}
\end{equation}

\noindent
This equation has two roots. The first one $v_\mathrm{crit,1}$
is given by the usual condition
$\vec{g_\mathrm{eff}}= \vec{0}$, which implies the equality
$\Omega^2 R^{3}_\mathrm{eb}/(GM) =1$ at the equator. 
This corresponds to an equatorial critical velocity

\begin{equation}
v_\mathrm{crit, 1} = \Omega \; R_\mathrm{eb} = 
\left( \frac{2}{3} \frac{GM}{R_\mathrm{pb}} \right)^{\frac{1}{2}} \; .
\label{Ma314}
\end{equation}

\noindent
$R_\mathrm{eb}$  and $R_\mathrm{pb}$ are respectively the 
equatorial  and polar radius at  the critical velocity. We notice
that the critical velocity $v_\mathrm{crit, 1}$ is independent
on the Eddington factor. To this extent, this is in agreement with
\cite{Gla98}. The basic physical reason for this
independence is quite clear:
the radiative flux tends toward zero when the effective gravity
is zero, thus there is no effect of the radiative acceleration when
this occurs.

Equation~\ref{Ma313}  has a second root, which is given by the condition
$\Gamma_{\Omega}(\vartheta)$ = 1.
If we call the  Eddington ratio $\Gamma_\mathrm{max}$ the maximum
value  of $\kappa(\vartheta) L(P)/(4 \pi c GM)$ over the surface 
(in general at equator), \cite{mm6} has shown that
for  $\Gamma_\mathrm{max} < 0.639$, no value of $\Omega$ can lead to
$\Gamma_{\Omega}(\vartheta)$ = 1.
Physically this means that
when the star is sufficiently far from the Eddington limit, the effects of
rotation on the radiative equilibrium are not sufficient for
lowering the Eddington luminosity such that it may have an impact on the
value of the critical velocity.
In that case, Eq.~\ref{Ma313} has only 
one root given by the classical expression.

For a given large enough
 $\Gamma_\mathrm{max}$ (i.e. larger than 0.639), 
a second root is obtained given by

\begin{eqnarray}
v_\mathrm{crit, 2}^2 =
\frac{9}{4} \;v_\mathrm{crit, 1}^2 \; 
\frac{1-\Gamma}{V^{\prime}(\omega)} \; \frac{
R^{2}_\mathrm{e}(\omega)}{R^2_{\mathrm{pb}}}
\end{eqnarray}

The quantity $\omega$ is the fraction $\Omega/\Omega_\mathrm{c}$
 of the angular velocity at break--up.
The quantity $V^{\prime}(\omega)$
is the ratio of the actual volume of a star with rotation $\omega$ to 
the volume of a sphere  of radius   $R_\mathrm{pb}$.
$V^{\prime}(\omega)$ is  obtained by the integration of the
solutions of the surface equation for a given value of
the parameter $\omega$. 

For $\Gamma_\mathrm{max} > 0.639$, this second root is inferior
to the first one. Thus it is encountered first and is therefore the expression of the
critical velocity that has to be used.
\vskip 2mm
\noindent{\underline{The mass loss rates}}
\vskip 2mm
Due to the von-Zeipel theorem, the radiative flux, which is the driving force for
the stellar winds of massive hot stars, varies as a function of the colatitude. This effect,
when accounted for in the theory of the line driven wind theory, lead to an enhancement of the quantities of mass lost and to wind anisotropies. We shall describe in more details
these effects in the Section 1.3 below. 
\vskip 2mm
\noindent{\underline{Three remarks}}
\vskip 2mm
1) The classical critical angular velocity or the $\Omega$-limit (to distinguish it from the $\Omega\Gamma$-limit as defined by \cite{mm6}) in the frame of the Roche model is given by 
\begin{equation}
\Omega_{\rm crit}=\left({2 \over 3}\right)^{3 \over 2}\left({GM \over R^3_{\rm pb}}\right)^{1 \over 2},
\label{eq4}
\end{equation}
where $R_{\rm pb}$ is the polar radius when the surface rotates with the critical velocity. The critical velocity is given by
\begin{equation}
\upsilon_{\rm crit}=\left({2 \over 3}{GM \over R_{\rm pb}}\right)^{1 \over 2}.
\label{eq4b}
\end{equation}
Replacing $\Omega$ in Eq.~\ref{eq3} by $(\Omega/\Omega_{\rm crit}) \cdot \Omega_{\rm crit}$ and using Eq.~\ref{eq4}, one obtains a relation between $R_{\rm p}$, $f$ and $R_{\rm pb}$,
\begin{equation}
R_{\rm p}={3 \over 2} R_{\rm pb} \left({\Omega_{\rm crit} \over \Omega}\right)^{2/3} \left({2 (f-1)\over f^3} \right)^{1/3},
\label{eq5}
\end{equation}
Thus one has
\begin{equation}
{\Omega \over \Omega_{\rm crit}}=\left({3 \over 2}\right)^{3/2} \left({R_{\rm pb} \over R_{\rm p}}\right)^{3/2} \left({2 (f-1)\over f^3} \right)^{1/2}.
\label{eq6}
\end{equation}
With a good approximation (see below) one has that $R_{\rm pb}/ R_{\rm p} \simeq 1$ and therefore
\begin{equation}
{\Omega \over \Omega_{\rm crit}} \simeq \left({3 \over 2}\right)^{3/2} \left({2 (f-1)\over f^3} \right)^{1/2}.
\label{eq7}
\end{equation}
This equation is quite useful since it allows the determination of $\Omega_{\rm crit}$ from quantities obtained with a model computed for $\Omega$. This is not the case of Eq.~\ref{eq5} which involves $R_{\rm pb}$ whose knowledge can only be obtained by computing models at the critical limit. In general Eq.~\ref{eq7} gives a very good approximation of $\Omega_{\rm crit}$ (see \cite{EMM}
for a discussion of this point).

\begin{figure}[t]
  \resizebox{\hsize}{!}{\includegraphics[angle=-90]{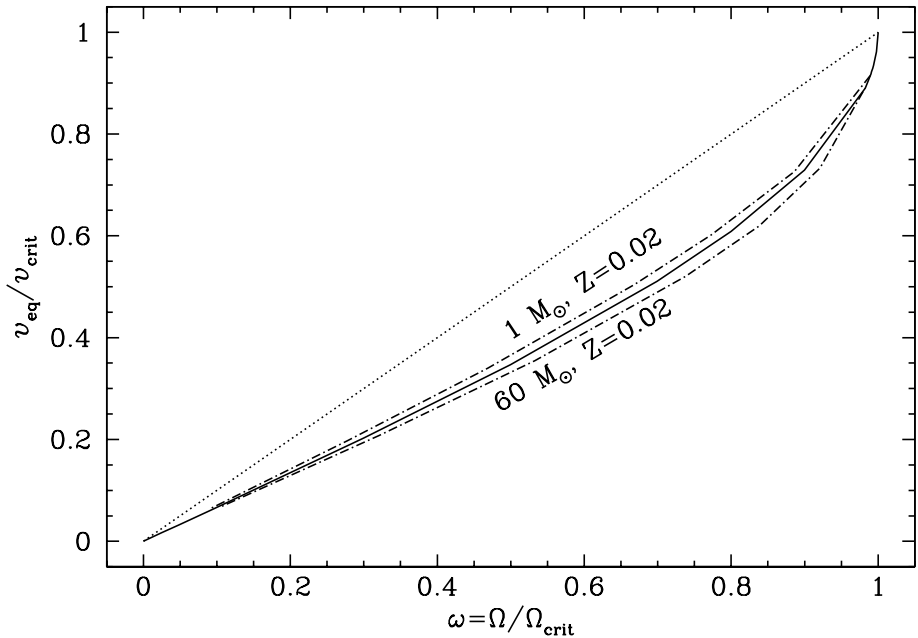}}
      \caption{Relation between $\upsilon/\upsilon_{\rm crit}$ and
      $\Omega/\Omega_{\rm crit}$ obtained in the frame of the Roche model. The
      continuous line is obtained assuming $R_{\rm pb}/R_{\rm p}=1$ (see text
      and Eqs.~\ref{eq7} and \ref{eq8}). The dot-dashed lines show the relations
      for the Z=0.02 models with 1 and 60 M$_{\odot}$ using Eq.~\ref{eq6}. The
      dotted line is the line of slope 1.}
       \label{ocvc}
  \end{figure}

Setting $\upsilon$ the velocity at the equator, one has that
\begin{equation}
{\upsilon \over \upsilon_{\rm crit}}={\Omega R_{\rm e}\over\Omega_{\rm c}R_{\rm eb}}
={\Omega \over \Omega_{\rm crit}}{R_{\rm e}\over R_{\rm p}}{R_{\rm p}\over R_{\rm pb}}{R_{\rm pb}\over R_{\rm eb}},
\label{eq7bis}
\end{equation}
where $R_{\rm eb}$ is the equatorial radius when the surface rotates with the critical velocity.
Using Eq.~\ref{eq5} above, and the fact that in the Roche model $R_{\rm pb}/ R_{\rm eb}=2/3$,
one obtains
\begin{equation}
{\upsilon \over \upsilon_{\rm crit}}=
\left({\Omega \over \Omega_{\rm crit}}2(f-1)\right)^{1/3}
\label{eq8}
\end{equation}
The relations between $\upsilon/\upsilon_{\rm crit}$ and $\Omega/\Omega_{\rm crit}$ obtained in the frame of the Roche model (see Eq.~\ref{eq8}) for the 1 and 60 M$_\odot$ stellar models at $Z=0.02$ are shown in Fig.~\ref{ocvc}. In case we suppose that the polar radius remains constant $R_{\rm pb}/R_{\rm p}=1$, then  Eq.~\ref{eq7} can be used and one obtains a unique relation between $\Omega/\Omega_{\rm crit}$ and $\upsilon/\upsilon_{\rm crit}$, independent of the mass, metallicity and evolutionary stage considered. One sees that the values of $\upsilon/\upsilon_{\rm crit}$ is smaller than that of $\Omega/\Omega_{\rm crit}$ by
at most $\sim$25\%. At the two extremes the ratios are of course equal.  

\begin{figure}[t]
\includegraphics[angle=00,width=8.8cm]{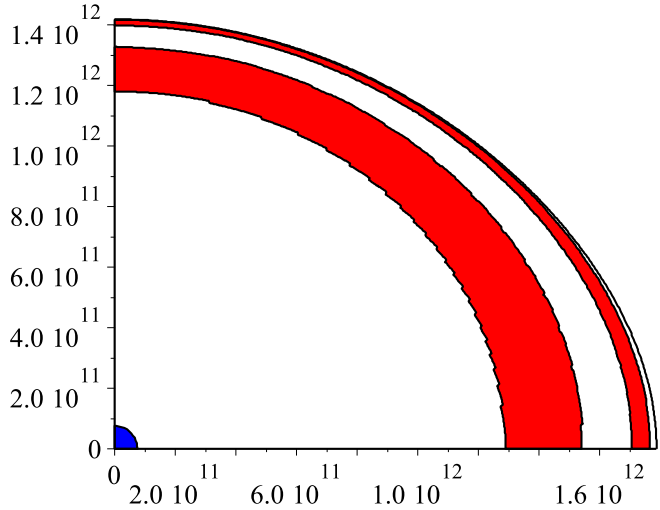}
\caption{2--D representation of the  external convective zones (in red) and of the convective core (blue)  in a model of 20 M$_{\odot}$ with $X=0.70$ and $Z=0.020$ at the end of MS evolution with fast rotation ( $\Omega/\Omega_{\mathrm{crit}}=0.94$). The axis are in units of cm.
Figure taken from \cite{MCM08}.}
\label{P94}
\end{figure}

An interesting quantity is the ratio of the centrifugal acceleration, $a_{\rm cen}$, to the gravity, $g_{\rm e}$, at the equator
\begin{equation}
{a_{\rm cen} \over g_{\rm e}}={\Omega^2 R^3_{\rm e}\over G M}=
\left({\Omega \over \Omega_{\rm crit}}\right)^2 \left({2 \over 3}\right)^3 f^3
\left({R_{\rm p}\over R_{\rm pb}}\right)^3,
\label{eq9}
\end{equation}
where we have used Eq.~\ref{eq4} and divided/multiplied by $R_{\rm pb}^3$. Replacing $R_{\rm pb}/R_{\rm p}$ by its expression deduced from Eq.~\ref{eq5}, we obtain
\begin{equation}
{a_{\rm cen} \over g_{\rm e}}=2(f-1).
\label{eq10}
\end{equation}
We can check that at the critical limit, when $f=3/2$, then $a_{\rm cen}=g_{\rm e}$.
\vskip 2mm
2) It is interesting to note that the fact that the effective temperature varies as a function
of the colatitude on an isobaric surface does not necessarily imply that the temperature
varies as a function of the colatitude on an isobaric surface. For instance in the conservative case, the temperature is constant on equipotentials which are also isobaric surfaces. This simply illustrates the difference between the effective temperature whose definition is related to the radiative flux ($F=\sigma T_{\rm eff}^4$) and hence to the {\it temperature gradient} and the temperature itself.
Interestingly, one has that in a conservative case, $\Gamma_\Omega(\theta)$ is constant on isobaric surfaces (neglecting
the term $\zeta(\vartheta)$). This means that when the $\Omega\Gamma$-limit is reached, it is reached
over the whole stellar surface at the same time. This is in contrast with the $\Omega$-limit which is reached
first at the equator.
\vskip 2mm
3) Recently we have examined the effects of rotation on the thermal gradient and on the Solberg--Hoiland term by analytical developments and by numerical models \cite{MCM08}. Writing the criterion for convection in rotating envelopes, we show that the effects of rotation on the thermal gradient are much larger and of  opposite sign to   the effect of the Solberg--Hoiland criterion. On the whole, rotation favors convection in stellar  envelopes at the  equator  and to a smaller extent at the poles. In a rotating 20 M$_{\odot}$ star at 94\% of the critical angular velocity, there are two convective envelopes, the biggest one has a thickness of 13.2\% of the equatorial radius.
The convective  layers are  shown in  Fig. \ref{P94}.  They are more extended  
than without rotation. 
In the non-rotating model, the corresponding convective zone has a thickness of only 4.6\% of the radius. The occurrence of outer convection in massive stars has many consequences (see \cite{MCM08}).

\vskip 2mm
\subsection{Transport mechanisms of angular momentum and of chemical species}

In a solid body rotation state, the first instability to set up is a thermal instability called the meridional circulation (see below). It consists in large meridional currents which transport angular momentum either inside-out or conversely from the outer regions toward the inner ones. Such meridional currents rapidly build ut gradients of the angular momentum both in the ``horizontal direction'' ({\it i.e.} along isobaric surface) and in the vertical one. Along isobaric surface, any gradient of $\Omega$ triggers a strong horizontal turbulence. Indeed in that direction the instability can develop without having to overcome any stable density gradient. As a consequence any
gradient of $\Omega$ along isobaric surfaces is rapidly erased and the star settles into a ``shellular'' rotation state \cite{Z92}. This means that $\Omega$ can be considered as nearly constant along isobars.
In the vertical direction, where in a radiative zone, a stable density gradient counteracts any instability, the gradients of $\Omega$ are eroded on much longer timescales  (see below). 
The equations below describe the interactions of meridional currents and of shear instabilities in a state of shellular rotation \cite{Z92}.


\subsubsection{Meridional circulation.}

Meridional circulation is an essential mixing mechanism in rotating
stars and there is a considerable literature on the subject
(see ref. in \cite{Tass90}). 
The velocity of the meridional circulation in the case of shellular 
rotation was derived by \cite{Z92}. The velocity of meridional circulation is derived from the equation of energy conservation \cite{Mes53}
\begin{eqnarray}
\rho T\left[{\partial S \over \partial t}+({\bf e}_r \dot r +{\bf U})\cdot{\bf \nabla}S\right]={\rm div}(\chi{\bf \nabla}T)+\rho\epsilon-{\rm div}{\bf F}_h
\label{eqn9}
\end{eqnarray}
where $S$ is the entropy per unit mass,
$\chi$ the thermal conductivity, $\epsilon$ the rate of nuclear energy per unit mass and ${\bf F}_h$ the flux of thermal energy due to
horizontal turbulence. All the quantities are expanded linearly around their average on a level surface or isobar, using
Legendre Polynomials $P_2(\cos\theta)$.
For instance
$$T(P,\theta)=\bar T(P)+\tilde{T}P_2(\cos\theta).$$
Then  Eq.~\ref{eqn9} is linearized and an expression
for $U_2$ can be deduced \cite{Z92}. Using the same method 
\cite{MZ98} revised the expression for $U_2$ to account for expansion and contraction in non--stationary models. They also studied the effects of the 
$\mu$--gradients (mean molecular weight gradients), of the horizontal turbulence and considered a general equation of state. They
obtained 
\begin{eqnarray}
U_2(r)={P \over \bar \rho \bar g
C_P\bar T[\nabla_{\rm ad}-\nabla+(\varphi/\delta)\nabla_\mu ]}
\times\left[{L\over M_*}(E_\Omega+E_\mu)+{C_P \over \delta}{\partial \Theta \over \partial
t}\right],   
\label{Umer}
\end{eqnarray}
where $M_{\star}=M \left( 1 - \frac{\Omega^2}{2 \pi G \rho_{\rm{m}}}  \right)$
is the reduced mass and the other symbols have the same meaning as in \cite{Z92} and \cite{MZ98}
\footnote{$U_2$ is the same as $U$ in Eq.~\ref{urt}}.
The driving term in the square brackets in the second member
 is $E_{\Omega}$.  It behaves mainly like
$E_{\Omega}  \simeq  \frac{8}{3} \left[ 1 - \frac{{\Omega^2}}
{2\pi G\overline{\rho}}\right] \left( \frac{\Omega^2r^3}{GM}\right)$
The term $\overline{\rho}$ means the average
on the considered equipotential.
The term with the minus sign in the square bracket is the 
Gratton--\"{O}pik term, which becomes important in the outer layers
when the local density is small. This term  produces negative values of $U_2(r)$
(noted $U(r)$ from now), meaning that 
the circulation is going down along the polar axis and up 
in the equatorial plane. This makes an outward transport 
of angular momentum, while a positive $U(r)$ gives an inward transport.
At lower $Z$, the Gratton--\"{O}pik term is negligible, which contributes
to make larger $\Omega$--gradients in lower $Z$ stars. 

Recently \cite{Mat04} 
rederived the system of partial differential equations, which govern the transport
of angular momentum, heat and chemical elements. They 
expand the departure from spherical symmetry to higher order and include explicitly the differential rotation 
in latitude, to first order. Boundary conditions for the surface and at 
the frontiers between radiative and convective zones are
also explicitly given in this paper.



\subsubsection{Shellular rotation.}

The differential rotation which results from the evolution and transport of the
angular momentum
makes the stellar
interior highly turbulent. As explained above, the turbulence is  very anisotropic, with
a much  stronger geostrophic--like transport
in the horizontal direction than in the vertical one \cite{Z92},
where stabilisation is favoured by the stable density gradient.
This strong horizontal transport is characterized by a 
large diffusion coefficient $D_{\rm{h}}$. Various expressions have been proposed:
\begin{itemize}
\item The usual expression for the coefficient $\nu_{\mathrm{h}}$ of 
viscosity  due to horizontal turbulence and for the coefficient $D_{\mathrm{h}}$ of
horizontal diffusion, which is of the same order, is, 
according to \cite{Z92},
\begin{equation}
D_{\mathrm{h}} \simeq \nu_{\mathrm{h}} =
\frac{1}{c_{\mathrm{h}}} r \;|2V(r) - \alpha U(r)| \,\, ,
\label{Zahn92}
\end{equation}
where $r$ is the appropriately defined eulerian coordinate 
of the isobar \cite{MM97}. 
$V(r)$ is defined by $u_\theta(r,\theta)=V(r) {{\rm d}P_2(\cos\theta) \over {\rm d}r}$ where
$u_\theta$ is  the horizontal component
of the velocity of the meridional currents\footnote{$V(r)$ can be obtained from $U(r)$ see
Eq.~2.10 in \cite{Z92}}, 
$\alpha = \frac{1}{2} \frac{d \ln r^{2} \Omega}
{d \ln r}$ and
$c_{\mathrm{h}}$ is a constant of  order of unity or smaller. 
This equation was derived assuming  that the 
differential rotation on an isobaric surface  is small \cite{MZ98}.
\item \cite{Ma99} has derived an expression for the coefficient $D_{\mathrm{h}}$ of diffusion by
horizontal turbulence in rotating stars.
He has obtained
\begin{eqnarray}
D_{\mathrm{h}} \propto \; r \; \left(r
\overline{\Omega}(r) \; V \;
 \left[ 2 V - \alpha U \right]\right)^\frac{1}{3} \;.
\label{nuh}
\end{eqnarray}
This expression  can be written  in
the usual form  $\nu_{\mathrm{h}}= \frac{1}{3} \; l \cdot v$ for a viscosity, 
where the appropriate velocity $v$
is a geometric mean of  3 relevant velocities: 
a velocity $(2 V - \alpha U)$ as in Eq~\ref{Zahn92} by \cite{Z92}, 
the horizontal component $V$ of the 
meridional circulation, 
the average local rotational velocity $ r \overline{\Omega}(r)$. This 
rotational velocity is usually much larger than either $U(r)$ or $V(r)$, typically by
6 to 8 orders of a magnitude in an upper Main Sequence star rotating with the average
velocity. 
\item From torque measurements in the classical Couette-Tayler experiment \cite{RZ99},
\cite{Mati04} have found the following expression
\begin{eqnarray}
\nu_{\mathrm{h}} = \left({\beta \over 10}\right)^{1/2}  \; \left(r^2
\overline{\Omega}(r) \; 
 \left[r \left|2 V - \alpha U \right|\right]\right)^\frac{1}{2} \; ,
\label{mat}
\end{eqnarray}
with $\beta\approx 1.5 \times 10^{-5}$ \cite{RZ99}.
\end{itemize}
The horizontal turbulent coupling  favours an essentially 
constant angular velocity $\Omega$   on the isobars. 
This rotation law, constant on shells,  applies to fast as well as to slow
rotators. As an approximation, it is often represented by a law of 
the form $\Omega = \Omega (r)$ (\cite{Z92}; see also \cite{ES76}).
Let us note here that the exact value of the diffusion coefficient $D_h$ is not well known.
Indeed the values of the numerical factors intervening in the various expressions shown
above may vary to some extent. Since the expression of $D_h$ intervenes in the formulas
for $U_r$, for $D_{\rm shear}$ and $D_{\rm eff}$ (see below), these uncertainties have some
impact on the amplitudes of the transport mechanisms. 

\subsubsection{Shear turbulence and mixing.}
In a radiative zone, shear due to differential rotation is likely
to be a most efficient mixing process. Indeed shear instability 
grows on a dynamical
timescale that is of the order of the rotation period 
\cite{Z92}.  
The usual criterion for shear instability
is the Richardson criterion, which compares the balance
between the restoring force of the density gradient
and the excess energy present in the differentially rotating layers,

\begin{equation}
Ri = \frac{N^{2}_{\mathrm{ad}}}{(0.8836\ \Omega\frac {d\ln\Omega}{d\ln r})^2} < \frac {1}{4},
\end{equation}

\noindent
where we have taken the average over an isobar,
$r$ is the radius and
$N_{\rm{ad}}$ the Brunt-V\"ais\"al\"a frequency given by

\begin{eqnarray}
N^2_{\rm{ad}} = \frac{g \delta}{H_{P}} \left[ \frac{\varphi}
{\delta} \nabla_{\mu} + \nabla_{ad} - \nabla_{\rm{rad}} \right].
\end{eqnarray}

\noindent
When thermal dissipation is 
significant, the restoring force of buoyancy is reduced and
the instability occurs more easily.
Its timescale is however longer, being the thermal timescale.
This case is  referred to as ``secular shear
instability''.  The criterion for low Peclet numbers $Pe$
(i.e. of large thermal dissipation, see below) has been 
considered by \cite{Za74}, while the cases of
general Peclet numbers  $Pe$ have been considered by
\cite{Mae95}, \cite{MM96}, 
who give 

\begin{equation}
Ri = \frac{g \delta}{(0.8836\ \Omega\frac {d\ln\Omega}{d\ln r})^{2} H_{P}} \left[
\frac{\Gamma}{\Gamma +1} (\nabla_{ad} -\nabla) +
\frac{\varphi}{\delta} \nabla_{\mu} \right] < \frac{1}{4}
\end{equation}

\noindent
The quantity $\Gamma =  Pe/6 $, where the Peclet number $Pe$
is the ratio of the thermal cooling time to the 
dynamical time, i.e.  $Pe = \frac{v \ell}{K}$ 
where $v$ and $\ell$ are the 
characteristic velocity and  length scales, and $K = (4acT^3)/
(3 C_P \kappa \rho^2 )$ is the thermal diffusivity. A discussion
of shear--driven turbulence by
\cite{Ca98} suggests that the limiting   $Ri$ number may
be larger than $\frac{1}{4}$. 

To account for shear transport and diffusion,
we need a diffusion coefficient. Amazingly, a great variety of
coefficients  $D_{\mathrm{shear}} = \frac{1}{3} v \ell$
 have been derived and applied (see a more extended discussion in \cite{MMV}):

\begin{enumerate} 
\item \cite{Z92} defines the diffusion coefficient cor\-responding
to the eddies which have the largest $Pe$ number so that the
Richardson criterion is just marginally satisfied. However, the
effects of the vertical $\mu$--gradient are not accounted for and
 the expression only applies to low Peclet numbers.
The same has been done by \cite{MM96},
who considers also the effect of the vertical $\mu$--gradient, the case of general
Peclet numbers and, 
in addition they account for the coupling due to the
fact that the shear also modifies the local thermal gradient.
This coefficient has been used by \cite{MM97}
and by \cite{Den99}.
The comparisons  of model results and observations of
surface abundances have led many authors to conclude that the 
$\mu$--gradients  appear to inhibit  the shear mixing
too much with respect to what is required by the observations
(\cite{Cha95a},\cite{MM97},\cite{He00}).

\item Instead of
using a gradient $\nabla_{\mu}$ in the criterion for shear mixing,
\cite{Cha95a} and \cite{He00}
 write  $f_{\mu} \nabla_{\mu}$ with a factor $f_{\mu} = 0.05$ 
or even  smaller.
This procedure is not satisfactory since it only accounts for a small fraction
of the existing $\mu$--gradients in stars.
The problem is that the models 
depend at least as much (if not more) on $f_{\mu}$ than on rotation, {\it i.e.} a change
of $f_{\mu}$ in the allowed range (between 0 and 1) produces as important effects
as a change of the initial rotational velocity.
This situation has led to two other more physical approaches discussed
below. Also \cite{He00} introduces another factor $f_c$ to adjust
the ratio of the transport of the angular momentum and of the
chemical elements like  \cite{Pin89}.

\item Around the convective core in the region
 where the $\mu$--gradient inhibits mixing,
there is anyway some turbulence due to both the  horizontal
turbulence and to the  semiconvective 
instability, which is  generally present in massive stars. 
This situation
has led  to the hypothesis  \cite{Mae97}
that the  excess energy in the shear,
or a fraction $\alpha$ of it of the order of unity, 
is degraded by turbulence on the local thermal timescale.
This progressively changes the entropy gradient 
and consequently the $\mu$--gradient. This hypothesis
leads to  a diffusion coefficient $D_{\rm shear}$ given by
\begin{eqnarray}
D_{\rm shear} = 4 \frac{K}{N^{2}_{\rm{ad}}} \left[ 
\frac{1}{4} \alpha \left(0.8836\  \Omega \frac{d\ln\Omega}{d\ln r}
\right)^2 - (\nabla^{\prime} -\nabla) \right].
\label{Mae97}
\end{eqnarray}
The term $\nabla^{\prime} -\nabla$ in Eq.~\ref{Mae97} expresses either
the stabilizing effect of the thermal gradients in radiative zones or
its destabilizing effect in semiconvective zones (if any). 
When the shear is negligible, 
$D_{\rm shear}$ tends toward the diffusion coefficient for semiconvection
by \cite{La83} in semiconvective zones.
When the thermal losses are large ($\nabla^{\prime} =\nabla$), it tends toward the value
\begin{equation}
D_{\rm shear} = \alpha (K/N^2_{\rm{ad}}) \left(0.8836\  \Omega \frac{d\ln\Omega}{d\ln r}\right)^2 ,
\label{orig}
\end{equation}
given by \cite{Z92}.
Eq.~\ref{Mae97} is completed by 
the three following equations expressing the thermal effects 
\cite{Mae97}
\begin{eqnarray}
D_{\rm shear} = 2 K \Gamma \;\;\;\;\;\;\;\; \nabla=\frac{\nabla_{rad}+
(\frac{6 \Gamma^2}{1+\Gamma}) \nabla_{ad}}{1+(\frac{6 \Gamma^2}{1+\Gamma})},
\label{Mae972}
\end{eqnarray}
\begin{eqnarray}
\nabla^{\prime} -\nabla = \frac{\Gamma}{\Gamma +1} (\nabla_
{\mathrm{ad}} - \nabla).
\label{Mae973}
\end{eqnarray}
The system of 4 equations given by Eqs.~\ref{Mae97}, \ref{Mae972} and \ref{Mae973} 
form a coupled
 system  with 4 unknown quantities
$D_{\rm shear}$, $\Gamma$, $\nabla$ and $\nabla^{\prime}$. 
The system is of the third degree in $\Gamma$.  
When it is solved   numerically,
we find  that as a matter of fact
the thermal losses in the shears are rather large in massive
stars  and thus that the Peclet number $Pe$ is very small
(of the order of 10$^{-3}$  to 10$^{-4}$).
For very low Peclet number
 $Pe =6 \Gamma$, the differences $(\nabla^{\prime} -\nabla)$ are 
also very small as shown by Eq.~\ref{Mae973}.
Thus, we conclude that Eq.~\ref{Mae97} is essentially equivalent,
at least in massive stars, to the original Eq.~\ref{orig} above, as
given by \cite{Z92}. We may suspect that this is not
 necessarily true in low and intermediate mass stars since there the $Pe$ number may
be larger. 

\item \cite{TZ97} found
that the diffusion coefficient for the shears
is modified by the horizontal turbulence. The change can be 
an increase or a decrease of the diffusion coefficient depending 
on the various parameters, as discussed below. Thus,
we have

\begin{eqnarray}
D =  \frac{ (K + D_{\mathrm{h}})}
{\left[\frac{\varphi}{\delta} 
\nabla_{\mu}(1+\frac{K}{D_{\mathrm{h}}})+ (\nabla_{\mathrm{ad}}
-\nabla_{\mathrm{rad}}) \right] }\; \times 
\label{dsh}
\\[2mm] \nonumber
 \frac{H_{\mathrm{p}}}{g \delta} \; 
\left [ \alpha\left( 0.8836\Omega{d\ln \Omega \over d\ln r} \right)^2
-4 (\nabla^{\prime}  -\nabla) \right]
\end{eqnarray}

\noindent
where $D_{\mathrm{h}}$ is the coefficient of  horizontal diffusion 
(cf. \cite{Z92}).
We ignore here the thermal coupling effects
discussed by Maeder (\cite{Mae97})
because they were found to be relatively small
 and  they increase the
numerical complexity. Interestingly, we see that
in regions where $\nabla_{\mu} \simeq 0$, Eq.~\ref{dsh}  leads us to replace 
$K$ by  $(K+D_{\mathrm{h}})$ in the usual expression 
(cf. \cite{TZ97}), i.e. it reinforces slightly
the diffusion in regions which are close to chemical homogeneity.
On the contrary, in regions where $\nabla_{\mu}$ dominates
with respect to $(\nabla_{\mathrm{ad}} -\nabla_{\mathrm{rad}})$,
 the transport
is proportional to $D_{\mathrm{h}}$ rather than to $K$,
 which is quite logical since the diffusion is then
determined by  $D_{\mathrm{h}}$ rather than by thermal effects.
The above result shows the importance of the
treatment for the meridional circulation, since in turn it
determines the size of  $D_{\mathrm{h}}$ and to some extent
the diffusion by shears. 
\end{enumerate}
Of course, 
the Reynolds condition 
$D_{\rm shear} \geq \frac{1}{3} \nu Re_c$ 
must be satisfied in order that  the medium is turbulent. The quantity
$\nu$ is the  total viscosity (radiative + mole\-cular) and 
 $Re_c$  the critical Reynolds number estimated to be
around 10 (cf. \cite{Den99}; \cite{Z92}).
 The numerical results indicate that the conditions
 for the occurrence of turbulence are satisfied.

\subsubsection{Transport of the angular momentum}

\begin{figure}[t]
\begin{center}
  \resizebox{5cm}{!}{\includegraphics[angle=0]{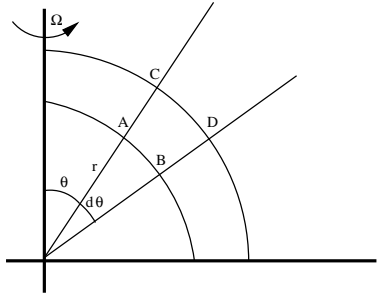}}
  \caption{The momentum of the viscosity forces acting on the element ABCD is derived in the text and
the general form of the equation describing the change with time of the angular momentum of this element
is deduced. The star rotates around the vertical axis with the angular velocity $\Omega$; $r$ and $\theta$ are
the radial and colatitude coordinates of point A.
}
  \label{schema}
\end{center}  
\end{figure}

Let us express the rate of change of the angular momentum, ${{\rm d}{\mathcal L}\over{\rm d}t}$, of the element of mass in the volume ABCD represented in Fig.~\ref{schema}:
$${{\rm d}{\mathcal L}\over{\rm d}t}={\bf M},$$
where ${\bf M}$ is the momentum of the forces acting on the volume element. 
We assume that angular momentum is transported only through advection (by a velocity field {\bf U}) and through turbulent diffusion, which may
be different in the radial (vertical) and tangential (horizontal) direction.
The component of the angular momentum aligned with the rotational axis is equal to
\footnote{The components perpendicular to the rotational axis cancel each other when the integration is performed over $\varphi$.} 
$$\underbrace{\rho r^2 \sin\theta {\rm d}\theta {\rm d}\varphi {\rm d} r}_{\rm Mass\ of\ ABCD}\ \ 
\underbrace{r\sin\theta \Omega}_{\rm velocity}\underbrace{r\sin\theta ,}_{\rm distance\ to\  axis}$$ 
where $ \Omega=\dot\varphi$.
Since the mass of the volume element ABCD does not change, the rate of change of the angular momentum can be written
\begin{eqnarray}
\rho r^2 \sin\theta {\rm d}\theta {\rm d}\varphi {\rm d} r {{\rm d}\over {\rm d}t} (r^2 \sin^2\theta \Omega)_{M_r}.
\label{eqn1}
\end{eqnarray}
Due to shear, forces apply on the surfaces of the volume element. The force on the surface AB
is equal to 
$$\underbrace{\eta_v}_{\rm vertical\  viscosity}\ \ \underbrace{r \sin\theta {\partial\Omega\over\partial r}}_{\rm vertical\ shear} 
\underbrace{r^2\sin\theta{\rm d}\theta{\rm d}\varphi}_{\rm surface\  AB}.$$ 
The component of the momentum of this force along the rotational axis is
$$\underbrace{\eta_v r^3\sin^2\theta {\partial\Omega\over\partial r}{\rm d}\theta{\rm d}\varphi}_{\rm force\  on\  AB} 
\ \underbrace{r\sin\theta .}_{\rm distance\  to\  axis}$$
The component along the rotational axis of the resultant momentum of the forces acting on AB and CD is equal to
\begin{eqnarray}
{\partial \over \partial r}(\eta_v r^4 \sin^3\theta {\rm d}\theta{\rm d}{\varphi}{\partial\Omega\over\partial r}){\rm d}r.
\label{eqn2}
\end{eqnarray}
The force on the surface AC due to the tangential shear is equal to
$$\eta_h \underbrace{r\sin\theta {\partial \Omega\over r\partial\theta}}_{\rm tangential\  shear}
\underbrace{r\sin\theta{\rm d}\varphi{\rm d}r }_{\rm surface\  AC},$$
where $\eta_h$ is the horizontal viscosity. The component along the rotational axis of the resultant momentum of the forces acting
on AC and BD is equal to
\begin{eqnarray}
{\partial \over r \partial \theta}(\eta_h r^2\sin^3\theta{\rm d}r{\rm d}\varphi {\partial \Omega \over \partial \theta})r{\rm d}\theta.
\label{eqn3}
\end{eqnarray}
Using Eqs.~\ref{eqn1}, \ref{eqn2} and \ref{eqn3}, simplifying by ${\rm d}r{\rm d}\theta{\rm d}\varphi$, one obtains the equation for the transport of the angular momentum
\begin{eqnarray}
\rho r^2 \sin\theta {{\rm d}\over {\rm d}t} (r^2 \sin^2\theta \Omega)_{M_r}=
{\partial \over \partial r}(\eta_v r^4 \sin^3\theta {\partial\Omega\over\partial r})+
{\partial \over \partial \theta}(\eta_h r^2\sin^3\theta {\partial \Omega \over \partial \theta}).
\label{eqn4}
\end{eqnarray}
Setting $\eta_v=\rho D_v$ and $\eta_h=\rho D_h$ and dividing the left and right member by $r^2\sin\theta$, one obtains
\begin{eqnarray}
\rho{{\rm d}\over {\rm d}t} (r^2 \sin^2\theta \Omega)_{M_r}={\sin^2\theta \over r^2} {\partial \over \partial r}(\rho D_v r^4  {\partial\Omega\over\partial r})+
{1\over \sin\theta}
{\partial \over \partial \theta}(\rho D_h \sin^3\theta {\partial \Omega \over \partial \theta}).
\label{eqn43}
\end{eqnarray}
Now, the left--handside term can be written
$$\rho{{\rm d}\over {\rm d}t} (r^2 \sin^2\theta \Omega)_{M_r}={{\rm d}\over {\rm d}t} (\rho r^2 \sin^2\theta \Omega)_{M_r}-r^2 \sin^2\theta \Omega{{\rm d}\rho\over {\rm
d}t}|_{M_r}.$$
Using the relation between the Lagrangian and Eulerian derivatives, one has
\begin{eqnarray}
\rho{{\rm d}\over {\rm d}t} (r^2 \sin^2\theta \Omega)_{M_r}=\nonumber
\end{eqnarray}
\begin{eqnarray}
{\partial\over \partial t} (\rho r^2 \sin^2\theta \Omega)_{r}
+{\bf U}\cdot{\bf \nabla}(\rho r^2 \sin^2\theta \Omega)-r^2 \sin^2\theta \Omega{{\rm d}\rho\over {\rm
d}t}|_{M_r}.
\label{eqn42}
\end{eqnarray}

\noindent Using
$${{\rm d} \rho \over {\rm d}t}|_{M_r}={\partial \rho \over \partial t}|_{r}+{\bf U} \cdot {\bf \nabla} \rho,$$
and the continuity equation
$${\partial \rho \over \partial t}|_{r}=-{\rm div}(\rho {\bf U}),$$
one obtains ${\rm d}\rho/{\rm d}t|_{M_r}+\rho {\rm div}{\bf U}=0$, which incorporated in Eq.~\ref{eqn42} gives
$$\rho{{\rm d}\over {\rm d}t} (r^2 \sin^2\theta \Omega)_{M_r}
={\partial \over \partial t} (\rho r^2 \sin^2\theta \Omega)_{r}+{\bf \nabla}({\bf U}\rho r^2 \sin^2\theta \Omega).$$
Developing the divergence in spherical coordinates and using Eq.~\ref{eqn43}, one finally obtains 
the equation describing the transport of the angular momentum (\cite{MZ98}; \cite{Mat04})
\begin{eqnarray}
{\partial\over \partial t} (\rho r^2 \sin^2\theta \Omega)_{r}+{1 \over r^2}{\partial \over \partial r}(\rho r^4\sin^2\theta w_r\Omega)+{1 \over r\sin\theta}
{\partial \over \partial \theta}(\rho r^2\sin^3\theta w_{\theta} \Omega)= \nonumber \\ {\sin^2\theta \over r^2} {\partial \over \partial r}(\rho D_v r^4  {\partial\Omega\over\partial r})+
{1\over \sin\theta}
{\partial \over \partial \theta}(\rho D_h \sin^3\theta {\partial \Omega \over \partial \theta}),
\label{eqn5}
\end{eqnarray}
where $w_r=U_r+\dot r$ is the sum of the radial component of the meridional circulation velocity and the velocity of expansion/contraction, and $w_\theta=U_\theta$,
where $U_\theta$ is the horizontal component of the meridional circulation velocity.
Assuming, as in \cite{Z92}, that the rotation depends little on latitude due to strong horizontal diffusion, we write
$$\Omega(r,\theta)=\bar\Omega(r)+\hat \Omega(r,\theta),$$
with $\hat\Omega \ll \bar\Omega$. The horizontal average $\bar \Omega$ is defined as being the angular velocity of a shell rotating
like a solid body and having the same angular momentum as the considered actual shell. Thus 
$$\bar \Omega={\int \Omega \sin^3 \theta{\rm d} \theta \over \int \sin^3 \theta {\rm d}\theta}.$$
Any vector field whose Laplacian is nul can be decomposed in spherical harmonics. Thus, the meridional circulation velocity can be written
\cite{Mat04}
$${\bf U}=\underbrace{\sum_{l > 0} U_l (r) P_l (\cos\theta)}_{u_r} {\bf e}_r+ \underbrace{
\sum_{l > 0} V_l(r) {{\rm d} P_l (\cos\theta) \over {\rm d} \theta}}_{u_\theta}{\bf e}_\theta,$$
where ${\bf e_r}$ and ${\bf e_\theta}$ are unit vectors along the radial and colatitude directions respectively.
Multiplying Eq.~\ref{eqn5} by $\sin\theta{\rm d}\theta$ and integrating it over $\theta$ from 0 to $\pi$, one obtains
\cite{MZ98}
\begin{eqnarray}
{\partial \over \partial t}(\rho r^2 \bar \Omega)_r={1 \over 5 r^2}{\partial \over \partial r}(\rho r^4 \bar \Omega [U_2(r)-5\dot r])
+{1 \over r^2}{\partial \over \partial r}\left(\rho D_v r^4 {\partial \bar \Omega \over \partial r} \right).
\label{eqn6}
\end{eqnarray}
It is interesting to note that only the $l=2$ component of 
the circulation is able to advect a net amount of angular momentum. As explained in
\cite{Spie92} the higher order components do not contribute to the vertical transport of angular momentum.
Note also that the change in radius $\dot r$ of the given mass shell is included in Eq.~\ref{eqn6}, which is the Eulerian formulation of the
angular momentum transport equation. In its Lagrangian formulation, the variable $r$ is linked to $M_r$ through ${\rm d}M_r=4\pi r^2 \rho {\rm d}r$, and
the equation for the transport of the angular momentum can be written
\begin{eqnarray}
\rho{\partial \over \partial t}(r^2 \bar \Omega)_{M_r}={1 \over 5 r^2}{\partial \over \partial r}(\rho r^4 \bar \Omega U_2(r))
+{1 \over r^2}{\partial \over \partial r}\left(\rho D_v r^4 {\partial \bar \Omega \over \partial r} \right).
\label{eqn7}
\end{eqnarray}
The characteristic time associated to the transport of $\Omega$ by the circulation is \cite{Z92}
\begin{eqnarray}
t_\Omega\approx t_{KH} \left({\Omega^2 R \over g_s}\right)^{-1},
\label{eqn8}
\end{eqnarray}
where $g_s$ is the gravity at the surface and $t_{KH}$ the Kelvin--Helmholtz timescale, which is the characteristic timescale for the change
of $r$ in hydrostatic models. From Eq.~\ref{eqn8}, one sees that $t_\Omega$ is a few times $t_{KH}$, which itself is much shorter that
the Main Sequence lifetime. 



For shellular rotation, the equation of transport of angular
momentum in the vertical direction is in lagrangian coordinates
(cf. \cite{Z92}; \cite{MZ98})

\begin{eqnarray}
\lefteqn{\rho \frac{d}{d t}
\left( r^2 \Omega\right)_{M_r} = } \nonumber \\[2mm]
&& \frac{1}{5 r^2}  \frac{\partial}{\partial r}
\left(\rho r^4 \Omega U(r) \right)
  + \frac{1}{r^2} \frac{\partial}{\partial r}
\left(\rho D r^4 \frac{\partial \Omega}{\partial r} \right) .
\label{full}
\end{eqnarray} 

\noindent $\Omega(r)$ is the mean angular velocity at level $r$.
The vertical component $u(r,\theta)$ of the velocity of the meridional
circulation at a distance $r$ to the center and at a colatitude $\theta$ can
be written

\begin{eqnarray}
u(r,\theta)=U(r)P_2(\cos \theta),
\label{urt}
\end{eqnarray} 

\noindent where $P_2(\cos \theta)$ is the second Legendre polynomial. Only the radial term
$U(r)$ appears in Eq.~\ref{full}.
The quantity $D$ is  the  total diffusion coefficient representing
the various instabilities considered  and which transport
the angular momentum, namely  convection, 
semiconvection and  shear turbulence. 
As a matter of fact, a
very large diffusion coefficient as in convective regions
implies a rotation law which is not far from solid body
rotation. In this work, we take  
$D = D_{\rm{shear}}$ in
radiative zones, since as extra--convective mixing
we  consider shear mixing and meridional circulation.

In case the outward transport of the angular momentum by the shear
is compensated by an inward transport due to the meridional circulation,
we obtain the local conservation of the 
angular momentum. We call this solution the \emph{stationary solution}.
In this case, $U(r)$ is given by (cf. \cite{Z92})

\begin{equation}
U(r)= - \frac{5 D}{\Omega} \frac{\partial \Omega}{\partial r} \; .
\label{stati}
\end{equation}

\noindent
The full solution of Eq.~\ref{full} taking into account $U(r)$ 
and $D$ gives the 
\emph{non--stationary solution} of the problem. In this case, 
$\Omega (r)$ evolves as a result of the various transport processes,
according to their appropriate timescales, and in turn 
differential rotation influences the various above processes.
This  produces a feedback and,  thus, a self--consistent
solution for the evolution of $\Omega (r)$ has to be found.

\begin{figure}[tb]
  \resizebox{\hsize}{!}{\includegraphics{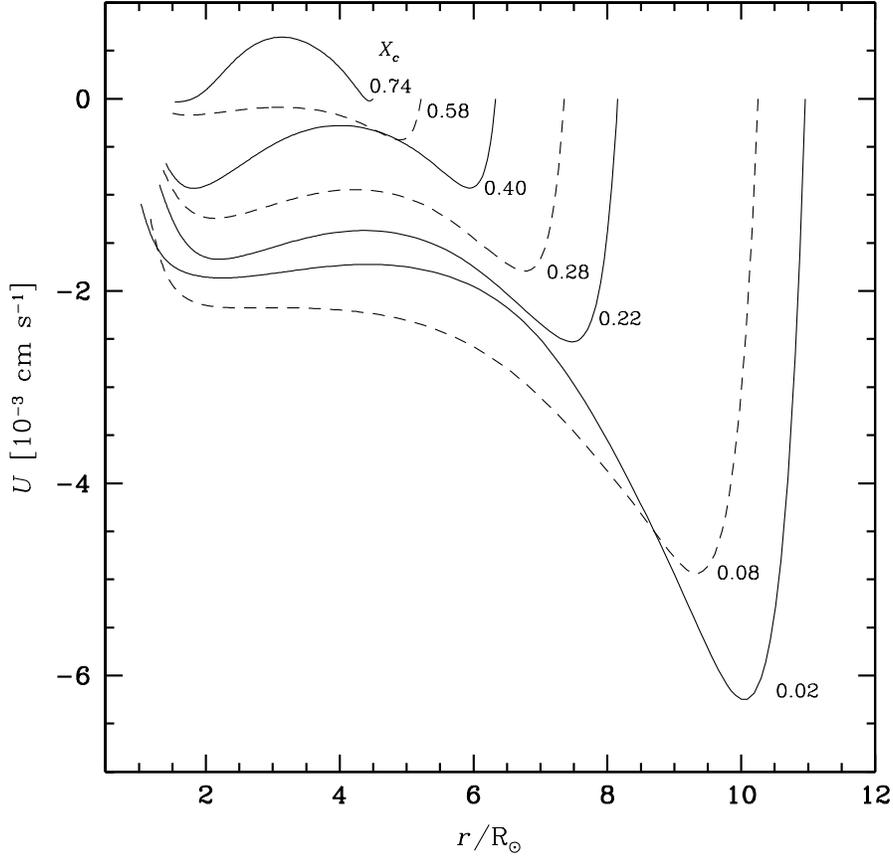}}
  \caption{Evolution of $U(r)$ the radial term of the vertical component of the
velocity of meridional circulation in a 20 M$_{\odot}$  star with $Z$ = 0.004 and an initial 
rotation velocity $v_{\mathrm{ini}}$ = 300 km s$^{-1}$. $X_c$ is the hydrogen mass 
fraction at the center. Figure taken from \cite{MMVII}}
  \label{Ur}
\end{figure}

Fig.~\ref{Ur} shows the evolution of $U(r)$  in a model
of a 20 M$_{\odot}$  star with $Z$ = 0.004 and an initial 
rotation velocity $v_{\mathrm{ini}}$ = 300 km s$^{-1}$ \cite{MMVII}.
 $U(r)$ is initially positive in the  interior, but
progressively  the 
fraction of the star where  $U(r)$ is negative is growing. This is
due to the Gratton--\"{O}pik term in Eq. (2),
which favors a negative $U(r)$ in the outer layers, 
when the density decreases. This negative velocity causes
 an outward transport of the angular momentum, as well as the
shears\footnote{When U is negative, the meridional currents turn anticlockwise, {\it i.e.} go
inwards along directions parallel to the rotational axis and go outwards in directions parallel to
the equatorial plane.}.

The transport of angular momentum by circulation 
has often been treated as a diffusion process
(\cite{ES76}; \cite{Pin89}; \cite{He00}). From Eq.~\ref{full}, we see
that the term with $U$ (advection)
is functio\-nally not the same as the term with $D$ (diffusion).
Physically advection and diffusion are quite different:
diffusion brings a quantity  from where there is a lot to other
places where there is little. This is not necessarily the case
for advection. A circulation with
a positive value of $U(r)$, i.e.\ rising 
along the polar axis and descending at the equator, 
is as a matter of fact making an inward transport
of angular momentum. Thus, we see that when this process is treated as a
diffusion, like a function of $\frac{\partial \Omega}{\partial r}$, even
the sign of the effect may be wrong.

The expression of $U(r)$ given above (Eq.~\ref{Umer}) involves derivatives up to the
third order, thus Eq.~\ref{full} is of the fourth order, which makes the
system very difficult to solve numerically. 
In practice, we have applied a Henyey scheme to make the
calculations. Eq.~\ref{full} also implies four boundary conditions.
At the stellar surface, we take (cf. 
\cite{Ta97})

\begin{eqnarray}
\frac{\partial \Omega}{\partial r} = 0  \;\;\;\;
\mathrm{and}  \;\;\; \; U(r) =0
\label{BE}
\end{eqnarray}

\noindent and at the edge of the core we have

\begin{eqnarray}
\frac{\partial \Omega}{\partial r} = 0  \; \; \; \;
\mathrm{and}  \; \; \; \;  \Omega(r) = \Omega_{\mathrm{core}.}
\end{eqnarray}

\noindent We  assume that 
the mass lost by stellar winds is just embarking its own angular momentum.
This means that we ignore any possible magnetic coupling, as it 
occurs in low mass stars.
It is interesting to mention here, that 
in case of no viscous, nor
magnetic coupling at the stellar surface, {\it i.e.} with the boundary
conditions \ref{BE},
the integration of Eq.~\ref{full} gives for an external shell of mass $\Delta M$ \cite{Ma99}

\begin{eqnarray}
\Delta M {d \over dt} (\Omega r^2)=-{4\pi \over 5} \rho r^4 \Omega U(r).
\end{eqnarray}

\noindent This equation is valid provided the stellar winds are spherically symmetric.
When the surface velocity approches the critical velocity, it is likely that there are  anisotropies of the mass
loss rates (polar ejection or formation of an equatorial ring)
and thus the surface condition should be modified according to 
the prescriptions of  \cite{Ma99}. 

\subsubsection{Mixing and transport of the chemical elements.}

A diffusion--advection equation like Eq.~\ref{full} should normally 
be used to express the 
transport of chemical elements. However, if the
horizontal  component of the turbulent diffusion $D_{\rm{h}}$
is large,
the vertical advection of the elements  can be treated as 
 a simple diffusion
\cite{Cha92} with a diffusion coefficient
$D_{\rm eff}$. As emphasized by \cite{Cha92} , this does
not apply to the transport of the angular momentum. $D_{\rm eff}$ is
given by

\begin{equation}
D_{\rm eff} = \frac{\mid rU(r) \mid^2}{30 D_h} \; ,
\label{deff}
\end{equation} 

\noindent where $D_{\rm{h}}$ is the coefficient of horizontal
turbulence. Eq.~\ref{deff}
expresses that the vertical
advection of chemical elements is severely inhibited by the
strong horizontal turbulence characterized by $D_{\rm{h}}$. 
 Thus, the change of the mass fraction $X_i$ of the chemical species 
$i$ is simply

\begin{eqnarray}
\left( \frac{dX_i}{dt} \right)_{M_r} =  
\left(\frac{\partial  }{\partial M_r} \right)_t
\left[ (4\pi r^2 \rho)^2 D_{\rm mix} \left( \frac{\partial X_i}
{\partial M_r}\right)_t
\right] +  \left(\frac{d X_i}{dt} \right)_{\rm nucl} .
\label{fullxi}
\end{eqnarray}

\noindent The second term on the right accounts for
 composition changes due
to nuclear reactions. The coefficient $D_{\rm mix}$ is the sum 
$D_{\rm mix} = D_{\rm shear}+D_{\rm eff}$
 and 
$D_{\rm eff}$ is given by Eq.~\ref{deff}. 
The characteristic time for the mixing
of chemical elements is therefore
$t_{\rm mix} \simeq \frac{R^2}{D_{\rm mix}} $
and is not given by  $t_{\rm circ} \simeq \frac{R}{U}$,
as has been generally considered \cite{Sch58}.
This makes the mixing of the chemical elements much slower,
since $D_{\rm eff}$ is very much reduced. 
In this context, we recall that several
authors have  reduced  by large factors, up to 30 or 100,
the coefficient for the transport
of the chemical elements, with respect to the
transport of the angular momentum, 
in order to better fit the observed 
surface compositions (cf. \cite{He00}). This
reduction of the diffusion of the chemical 
elements is no longer
necessary with the more appropriate expression of $D_{\rm eff}$ given here.

\begin{figure}
 \includegraphics[height=0.755\textheight]{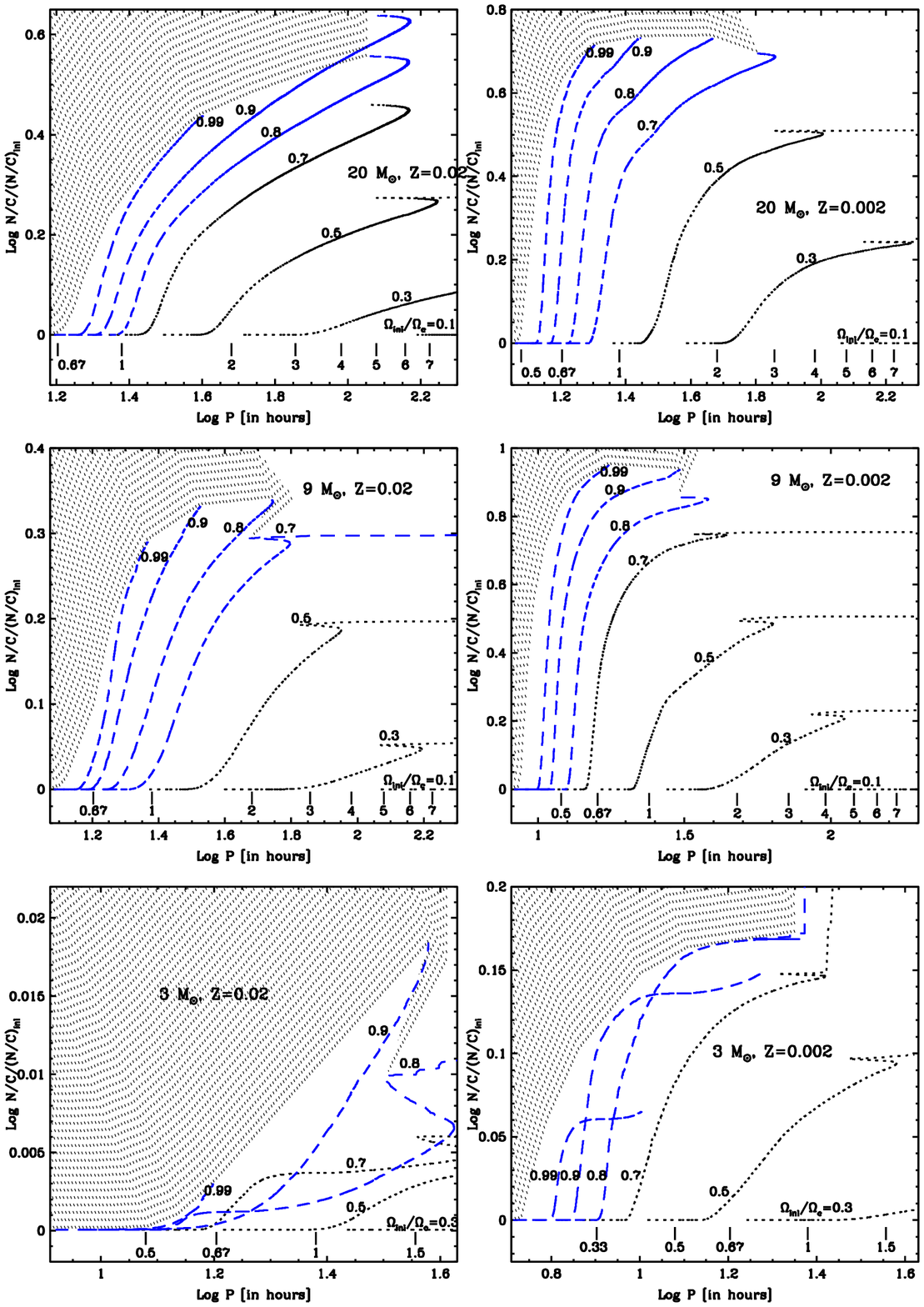}
      \caption{Evolutionary tracks in the plane surface N/C ratio,
      normalized to its initial value, versus the rotational
      period in hours for different initial mass stars, various initial
      velocities and for the metallicities $Z$=0.02 and 0.002. Positions of some
      periods in days are indicated at the bottom of the figure. The dotted
      tracks never reach the critical limit during the MS phase. The short
      dashed tracks reach the critical limit during the MS phase. The dividing
      line between the shaded and non-shaded areas corresponds to the entrance
      into the phase when the star is at the critical limit during the MS phase.
      If Be stars are stars rotating at or very near the critical limit, present
      models would predict that they would lie in the vicinity of this dividing
      line or above it. Note the different vertical scales used when comparing
      similar masses at different metallicities. Figure taken from \cite{EMM}.}
       \label{ncp}
  \end{figure}

Surface enrichments due to rotation are illustrated in Fig~\ref{ncp}.
The tracks are plotted in the plane ${\rm (N/C)/(N/C)_{ini}}$ versus $P$ where $P$ is the rotational period in hours. 
During the evolution the surface is progressively enriched in CNO burning
products, {\it i.e.} is enriched in nitrogen and depleted in carbon. At the same time, the rotational period  increases.

When the effects of the shear
and of the meridional circulation compensate each other for the
transport of the angular momentum (\emph{stationary solution}), the value of $U$
entering the expression for $D_{\rm eff}$ is given by Eq.~\ref{stati}.

\subsection{Rotation and mass loss}

We can classify the effects of rotation on mass loss in three categories.

\begin{enumerate}
\item The structural effects of rotation.
\item The changes brought by rotation on the radiation driven stellar winds.
\item The mass loss induced by rotation at the critical limit.
\end{enumerate}

Let us now consider in turn these various processes.

\subsubsection{Structural effects of rotation on mass loss.}

Rotation, by changing the chemical structure of the star, modifies
its evolution. For instance, moderate rotation at metallicities of the Small Magellanic Cloud (SMC)
favors redward evolution in the Hertzsprung-Russel diagram. This behavior can account for the high number of red supergiants observed in the SMC \cite{MMVII},
an observational fact which is not at all reproduced by non-rotating stellar models.

Now it is well known that the mass loss rates are greater
when the star evolves into the red part of the HR diagram, thus in this case, rotation modifies 
the mass loss indirectly, by changing the evolutionary tracks. 
The $\upsilon_{\rm ini}=0$, 200, 300 and 400 km s$^{-1}$ models lose respectively 0.14, 1.40, 1.71 and 1.93 M$_\odot$ during the core He-burning phase (see Table~1 in \cite{MMVII}). The enhancement of the mass lost reflects the longer lifetimes of the red supergiant phase when velocity increases.
Note that these numbers were obtained assuming that the same scaling law between mass loss and metallicity as in the MS phase
applies during the red supergiant phase. If, during this phase, mass loss comes from continuum-opacity driven
wind then the mass-loss rate will not depend on metallicity (see the review by \cite{vL06}).
In that case, the redward evolution favored by rotation would have a greater impact on mass loss than
that shown by the computations shown above.

Of course, such a trend cannot continue forever. For instance, 
at very high rotation, the star will have a homogeneous evolution and will never become a red supergiant
\cite{M87}.
In this case, the mass loss will be reduced, although this effect will be somewhat  compensated by
other processes: first by the fact that the Main-Sequence lifetime will last longer, second,
by the fact that the star will enter the Wolf-Rayet phase (a phase with high mass loss rates) at an earlier stage of its evolution, and third by the fact that the star may encounter the $\Omega$-limit.

\subsubsection{Radiation driven stellar winds with rotation.}

The effects of rotation on the radiation driven stellar winds 
result from the changes brought by rotation to the stellar surface. They induce changes of the morphologies
of the stellar winds and increase their intensities.
\vskip 2mm
\noindent\underline{Stellar wind anisotropies}
\vskip 2mm
Naively we would first guess that a rotating
      star would lose mass preferentially from the equator, where the effective gravity (gravity decreased
      by the effect of the centrifugal force) is lower.
      This is probably true when the star reaches the $\Omega$-limit (i.e. when the equatorial surface
      velocity is such that the centrifugal acceleration exactly compensates the gravity), but this is not
      correct when the star is not at the critical limit. Indeed as recalled above, a rotating star has a
      non uniform surface brightness, and the polar regions are those which have the most powerful radiative 
      flux. Thus one expects in case the opacity does not vary at the surface, that the star will lose mass preferentially along the rotational axis. This is
      correct for hot stars, for which the dominant source of opacity is electron scattering. In that
      case the opacity only depends on the mass fraction of hydrogen and does not depends on other
      physical quantities such as temperature. In that way, rotation induces 
      anisotropies of the winds   (\cite{MD01};\cite{DO02}).
      This is illustrated in the left panel of Fig.~\ref{ani}.
      Wind anisotropies have consequences for the angular momentum that a star retains in its interior.
      Indeed, when mass is lost preferentially along the polar axis, little angular momentum is lost.
      This process allows loss of mass without too much loss of angular momentum a process which might
      be important in the context of the evolutionary scenarios leading to Gamma Ray Bursts. Indeed 
      in the framework of the collapsar scenario (\cite{W93}), 
      one has to accommodate two contradictory requirements: on one side, the progenitor needs to lose mass
      in order to have its H and He-rich envelope removed at the time of its explosion, and on the other hand
      it must have retained sufficient angular momentum in its central region to give birth to a fast
      rotating black-hole.
      
\vskip 2mm      
\noindent\underline{Intensities of the stellar winds}      
\vskip 2mm      
      The quantity of mass lost through radiatively driven stellar winds is enhanced by rotation. This enhancement can occur through two channels: by reducing the effective gravity at the surface of the star, by
      increasing the opacity of the outer layers through surface metallicity enhancements due to rotational mixing.
      
\begin{itemize}

\item{\it reduction of the effective gravity: } The ratio of the mass loss rate of a star with a surface angular velocity $\Omega$ to that
      of a non-rotating star, of the same initial mass, metallicity and lying at the same position in the
      HR diagram is given by \cite{mm6}
      
      \begin{equation}
\frac{\dot{M} (\Omega)} {\dot{M} (0)} \simeq
\frac{\left( 1  -\Gamma\right)
^{\frac{1}{\alpha} - 1}}
{\left[ 1 - 
\frac{4}{9} (\frac{v}{v_{\mathrm{crit, 1}}})^2-\Gamma \right]
^{\frac{1}{\alpha} - 1}} \; ,
\end{equation}
\noindent
where $\Gamma$ is the electron scattering opacity for a non--rotating
star with the same mass and luminosity, $\alpha$ is a force multiplier \cite{La95}. 
The enhancement factor remains modest for stars with luminosity sufficiently far away from the
      Eddington limit \cite{mm6}. Typically, $\frac{\dot{M} (\Omega)} {\dot{M} (0)} \simeq 1.5$
      for main-sequence B--stars.
      In that case, when the surface velocity approaches the critical limit, the effective
      gravity decreases and the radiative flux also decreases. Thus the matter becomes less bound
      when, at the same time, the radiative forces become also weaker. 
      When the stellar luminosity approaches the Eddington limit, the mass loss increases can be much greater,
      reaching orders of magnitude.
This comes from the fact that rotation lowers the maximum luminosity or the Eddington luminosity of a star.  
      Thus it may happen that for a velocity still 
      far from the classical critical limit, the 
      rotationally decreased maximum luminosity becomes equal 
      to the actual luminosity of the star. 
      In that case, strong mass loss ensues and the star is said to have reached
      the $\Omega\Gamma$ limit \cite{mm6}.

\item {\it Effects due to rotational mixing: }
During the core helium burning phase, at low metallicity,
the surface may be strongly enriched in both H-burning and He-burning products, {\it i.e.} mainly in nitrogen, carbon and oxygen. Nitrogen is produced by transformation of the carbon and oxygen produced in the He-burning core and which have diffused by rotational mixing in the H-burning shell \cite{MMVIII}. Part of the carbon and oxygen produced in the He-core also diffuses up to the surface. Thus at the surface, one obtains very high value of the CNO elements. For instance a 60 M$_\odot$ with Z=$10^{-8}$ and $\upsilon_{\rm ini}=800$ km s$^{-1}$ has, at the end of its evolution, a CNO content at the surface equivalent to 1 million times its initial metallicity! In
case the usual scaling laws linking the surface metallicity
to the mass loss rates is applied, such a the star would lose due to this process
more than half of its initial mass.

\end{itemize}
      
\subsubsection{Mass loss induced by rotation}      
      
\begin{figure}
\resizebox{5cm}{!}{\includegraphics{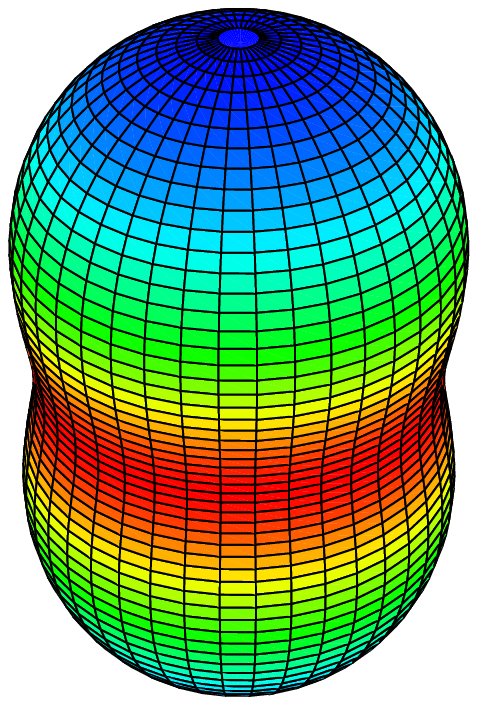}}
\hfill
\includegraphics[width=2.5in,height=2.5in]{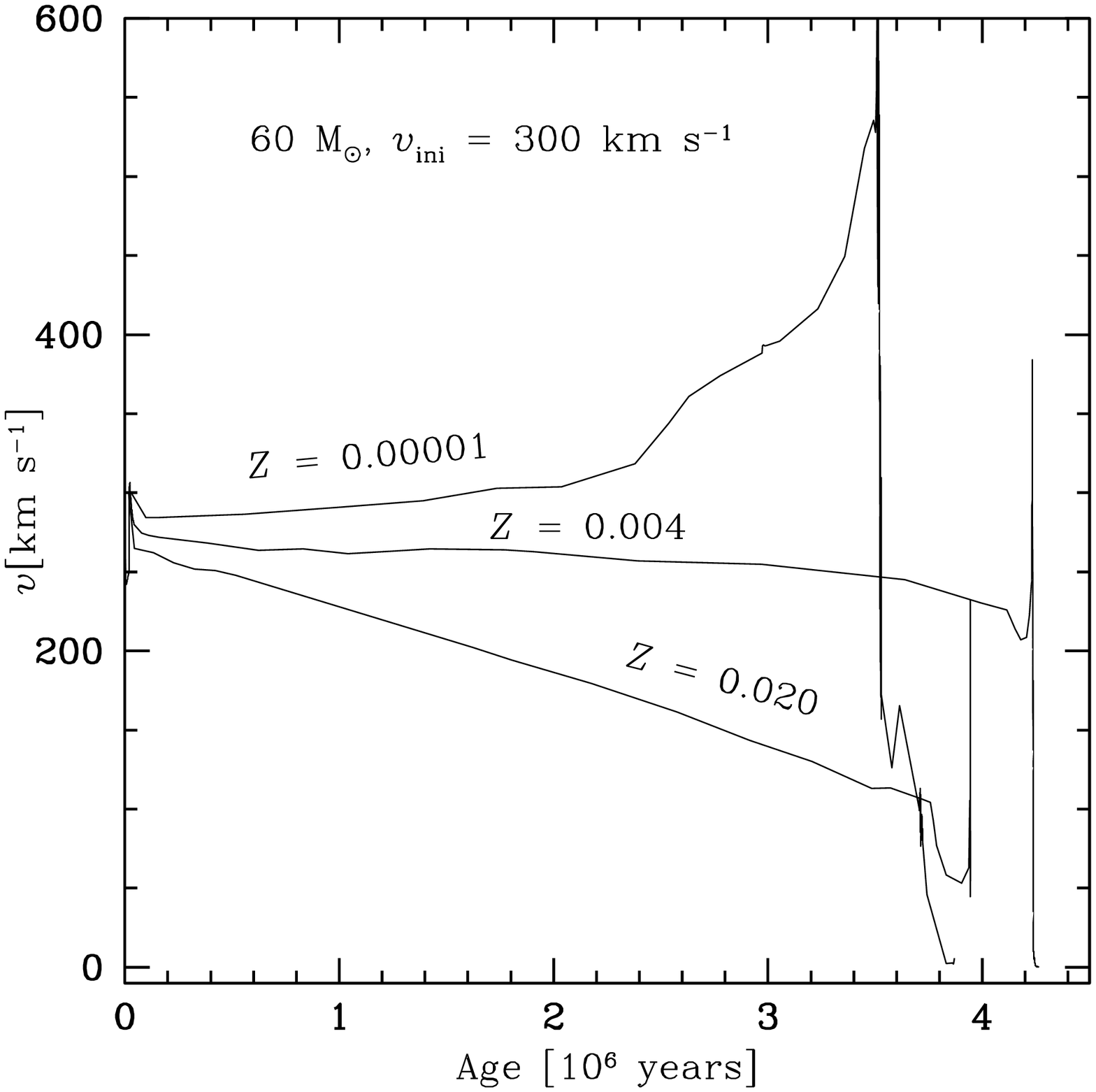}
\caption{{\it Left panel}: Iso-mass loss distribution for a 120 M$_\odot$ star with Log L/L$_\odot$=6.0 and T$_{\rm eff}$ = 30000 K rotating at a fraction 0.8 of critical velocity (figure
from \cite{MD01}).   
{\it Right panel}: Evolution of the surface velocities for a 60 M$_
 {\odot}$ star with 3 different
 initial metallicities. }
\label{ani}
\end{figure}

As recalled above, during the Main-Sequence phase the core contracts
      and the envelope expands. In case of local conservation of the angular momentum, the core would thus
      spin faster and faster while the envelope would slow down. In that case, it can be easily shown that the surface velocity would evolve away from the critical velocity (see e.g. \cite{Vl06}). 
      In models with shellular rotation however
      an important coupling between the core and the envelope is established through the action of the
      meridional currents. As a net result, angular momentum is brought from the inner regions to the outer ones. Thus, would the star lose no mass by radiation driven stellar winds (as is the case at low Z), one expects that the surface velocity
      would increase with time and would approach the critical limit. In contrast, 
      when radiation driven stellar winds are important, the timescale for removing mass 
      and angular momentum at the surface
      is shorter than the timescale for accelerating the outer layers by the above process and the surface
      velocity decreases as a function of time. It evolves away from the critical limit. 
      Thus, an interesting situation occurs: when the star loses
      little mass by radiation driven stellar winds, it has more chance to lose mass by reaching the critical limit. On the other hand, when the star loses mass at a high rate by
      radiation driven mass loss then it has no chance to reach the critical limit and thus to undergo a 
      mechanical mass loss. This is illustrated in the right panel of Fig.~\ref{ani}.
      
\subsubsection{Discussion.}  

At this point it is interesting to discuss three aspects of the various effects described above. First what
are the main uncertainties affecting them?  Second, what are their relative importance? And finally what are their consequences for the interstellar medium enrichment?
\vskip 2mm
\noindent\underline{Uncertainties}
\vskip 2mm
In addition to the usual uncertainties affecting the radiation driven mass loss rates, the above processes 
poses three additional problems:
\begin{enumerate}

\item {\it What does happen when the CNO content of the surface increases by six orders of magnitude as was obtained
in the 60 M$_\odot$ model described above?} Can we apply the usual scaling law between Z and the mass losses?
This is what we have done in our models, but of course this should be studied in more details by stellar winds models. For instance, for WR stars, \cite{V05} have shown that at $Z=Z_\odot/30$, 60\% of the driving is due to CNO elements and only 10\% to Fe.
Here the high CNO surface enhancements result
from rotational mixing which enrich the radiative outer region of the star in these elements, but also from the fact that the star evolves to the red part of the HR diagram, making an outer convective zone to appear. This
convective zone plays an essential role in dredging up the CNO elements at the surface. Thus what is needed
here is the effects on the stellar winds of CNO enhancements in a somewhat red part of the HR diagram
(typical effective temperatures of the order of Log T$_{\rm eff}\sim$3.8).

\item {\it Do stars can reach the critical limit?} For instance, \cite{BO70}
obtain that during pre-main sequence evolution of rapidly rotating massive stars, ``equatorial mass loss'' or ``rotational mass ejection'' never occur (see also \cite{BO73}). In these models the condition of zero effective gravity is never reached.  However, these authors studied pre-main sequence evolution and made different hypotheses on the
transport mechanisms than in the present work. Since they were interested in the radiative contraction
phase, they correctly supposed that ``the various instabilities and currents which transport angular momentum
have characteristic times much longer than the radiative-contraction time''. This is no longer the case 
for the Main-Sequence phase. In our models, we consistently accounted for the transport of the angular
momentum by the meridional currents and the shear instabilities.
A detailed account of the transport mechanisms shows that they are never able to prevent the star from reaching
the critical velocity.
Another difference between the approach in the work of \cite{BO70} and ours is that
\cite{BO70} consider another distribution of the angular velocity than in our models. They supposed constant $\Omega$
on cylindrical surface, while here we adopted, as imposed by the theory of \cite{Z92}, a ``shellular rotation law''. They resolved the Poisson equation for the gravitational potential, while here we adopted the Roche model. Let us note that the Roche approximation appears justified in the present case, since only the outer layers, containing little mass, are approaching the critical limit. The majority of the stellar mass has a rotation rate much below the critical limit and is thus not strongly deformed by rotation. Thus these differences probably explain why in our models we reach situations where the effective gravity becomes zero.

\item {\it What does happen when the surface velocity reaches the critical limit?} 
Let us first note that when the surface reaches the critical velocity, the energy which is still needed to
make equatorial matter to escape from the potential well of the star is still important. This is because the gravity of the system continues of course to be effective all along the path from the surface to the infinity
and needs to be overcome.
If one
estimates the escape velocity from the usual equation energy for a piece of material of mass $m$ at the equator
of a body of mass $M$, radius $R$ and rotating at the critical velocity,
\begin{equation}
{1 \over 2}m \upsilon_{\rm crit}^2+{1 \over 2}m \upsilon_{\rm esc}^2-{GMm \over R}=0,
\end{equation}
one obtains, using $\upsilon_{\rm crit}^2={GM/R}$  that the escape velocity is 
simply reduced by a factor $1/\sqrt{2}=0.71$ with respect to the escape velocity from a non-rotating body
\footnote{We suppose here that
the vector $\upsilon_{\rm esc}$ is normal to the direction of the vector $\upsilon_{\rm crit}$.}.
Thus the reduction is rather limited and one can wonder if matter will be really lost.
A way to overcome this difficulty is to consider the fact that, at the critical limit, the matter will
be launched into a keplerian orbit around the star.
Thus, probably, when the star reaches
the critical limit an equatorial disk is formed like for instance around Be stars. 
Here we suppose that this disk will eventually dissipate by radiative effects and thus that the material will be lost by the star.

Practically, in the present models, we remove the supercritical layers. This removal of material allows the outer layers to become again subcritical at least until secular evolution will bring again the surface near the critical limit (see \cite{MEM06} for more details
in this process). Secular evolution during the Main-Sequence phase triggers two counteracting effects: on one side, the stellar surface expands. Local conservation of the angular momentum makes the surface to slow down and the surface velocity to evolve away from the critical limit. On the other
hand, meridional circulation continuously brings angular momentum to the surface and accelerates the outer layers. This last effect in general overcomes the first one and the star rapidly reach again the critical limit. How much mass is lost by this process?
As seen above, the two above processes will maintain the star near the critical limit for most of the time.
In the models, we adopt the mass loss rate required 
to maintain the star at about 95-98\% of the critical limit. Such a mass loss rate is imposed  as long as the secular evolution brings back the star near the critical limit. In general, during the
Main-Sequence phase, once the critical limit is reached, the star remains near this limit for the rest
of the Main-Sequence phase. At the end of the Main-Sequence phase, evolution speeds up and
the local conservation of the angular momentum overcomes the effects due to meridional currents, the star
evolves away from the critical limit and the imposed ``critical'' mass loss is turned off.

\end{enumerate}

\section{``Spinstars'' at very low metallicities?} 

Let us call ``spinstars'' those stars with a sufficiently high initial rotation in order to have their evolution significantly affected by rotation. In this section, we present some arguments supporting the view according to which spinstars might have been more common in the first generations of stars in the Universe.
A direct way to test this hypothesis would be to obtain measures of surface velocity of very metal poor massive stars and to see whether their rotation is superior to those measured at solar metallicity. At the moment, such measures can be performed only for a narrow range of metallicities for $Z$ between 0.002
and 0.020. Interestingly already some effects can be seen. For instance \cite{Keller04} presents measurements of the projected rotational velocities of a sample of 100 early B-type main-sequence stars in the Large Magellanic Cloud (LMC). He obtains that the stars of the LMC are more rapid rotators than their Galactic counterparts and that, in both galaxies, the cluster population exhibits significantly more rapid rotation than that seen in the field (a point also recently obtained by \cite{HG06}). More recently \cite{Martayan07} obtain that the angular velocities of B (and Be stars) are higher in the SMC than in the LMC and MW. 

For B-type stars, the higher values obtained at lower $Z$ can be the result of two processes: 1) the process of star formation produces more rapid rotators at low metallicity; 2) the mass loss
being weaker at low $Z$, less angular momentum is removed from the surface and thus starting from the same initial velocity, the low $Z$ star would be less slowed down by the winds. In the case of B-type stars, the mass loss rates are however quite modest and we incline to favor the first hypothesis, {\it i.e.} a greater fraction of fast rotators at birth at low metallicity. Another piece of argument supporting this view is the following: in case the mass loss rates are weak (which is the case on the MS phase for B-type stars), then the surface velocity is mainly determined by two processes, the initial value on the ZAMS and the efficiency of the angular momentum transport from the core to the envelope. In case of very efficient transport, the surface will receive significant amount of angular momentum transported from the core to the envelope. The main mechanism responsible for the transport of the angular momentum is meridional circulation. The velocities of the meridional currents in the outer layers are smaller when the density is higher thus in more metal poor stars. Therefore, starting from the same initial velocity on the ZAMS, one would expect that B-type stars at solar metallicity (with weak mass loss) would have higher surface velocities than the corresponding stars at low $Z$. The opposite trend is observed. Thus, in order to account for the higher velocities of B-type stars in the SMC and LMC, in the frame of the present rotating stellar models, one has to suppose that stars on the ZAMS have higher velocities at low $Z$. Very interestingly, the fraction of Be stars (stars rotating near the critical velocity) with respect to the total number of B stars is higher at low metallicity (\cite{MaederGrebel99}; \cite{Wisniewski06}). This confirms the trend
discussed above favoring a  higher fraction of fast rotators at low $Z$.

There are at least four other striking observational facts which might receive an explanation based on massive fast rotating models. {\it First}, indirect observations indicate the presence of very helium-rich stars in the globular cluster $\omega$Cen \cite{Piotto05}. Stars with a mass fraction of helium, $Y$,  equal to 0.4 seem to exist, together with a population of normal helium stars with $Y=0.25$. 
Other globular clusters appear to host helium-rich stars \cite{Caloi07}, thus
the case of $\omega$Cen is the most spectacular but not the only one.
There is no way for these very low mass stars to enrich their surface in such large amounts of helium and thus they must have formed from protostellar cloud having such a high amount of helium. Where does this helium come from? We proposed that it was shed away by the winds of metal poor fast rotating stars \cite{MMocen}. 

{\it Second}, in globular clusters, stars made of material only enriched in H-burning products have been observed (see the review by \cite{Gratton04}). Probably these stars are also enriched in helium and thus this observation is related to the one reported just above. The difference is that proper abundance studies can be performed for carbon, nitrogen, oxygen, sodium, magnesium, lithium, fluorine \dots, while for helium only indirect inferences based on the photometry can be made. \cite{DecressinI}  propose that the matter from which the stars rich in H-burning products are formed, has been released by slow winds of fast rotating massive stars. Of course, part of the needed material can also be released by AGB stars. The massive star origin presents however some advantages: first a massive star can induce star formation in its surrounding, thus two effects, the enrichment and the star formation can be triggered by the same cause. Second, the massive star scenario allows to use a less flat IMF than the scenario invoking AGB stars \cite{PC06}. The slope of the IMF might be even a Salpeter's one in case the globular cluster lost a great part of its first generation stars by tidal stripping (see \cite{DCM07}).

{\it Third}, the recent observations of the surface abundances of very metal poor halo stars\footnote{These stars are in the field and present [Fe/H] as low as -4, thus well below the metallicities of the globular clusters.} show the need of a very efficient mechanism for the production of primary nitrogen \cite{Chiappinial05}. As explained in \cite{Chiappinial06}), a very nice way to explain this very efficient primary nitrogen production is to invoke fast rotating massive stars. Very interestingly, fast rotating massive stars help not only in explaining the behavior of the N/O ratio at low metallicity but also those of the C/O. Predictions for the behaviour of the $^{12}$C/$^{13}$C ratios at the surface of very metal poor non-evolved stars have also been obtained \cite{Chiappinial08}.

{\it Fourth}, below about [Fe/H] $<$ -2.5, a significant fraction of very iron-poor stars are C-rich
(see the review by \cite{BC05}). Some of these stars show no evidence of $s$-process enrichments by AGB stars and are thus likely formed from the ejecta of massive stars. The problem is how to explain the very high abundances with respect to iron of CNO elements.
\cite{MEM06} and \cite{Hi07} proposed that these stars might be formed from the winds of very metal poor fast rotating stars. It is likely that rotation also affects the composition of the ejecta of intermediate mass stars. \cite{MEM06} predict the chemical composition of the envelope of a 7 M$_\odot$ E-AGB star which have been enriched by rotational mixing. The composition presents striking similarities with the abundance patterns observed at the surface of CRUMPS. The presence of overabundances of fluorine and of $s$-process elements might be used to discriminate between massive and intermediate mass stars.

All the above observations seem to point toward the same direction, an important population of spinstars at low Z. How many? What is the origin of the fast rotation? What are the consequences for the Gamma ray Burst progenitors?
All these questions have still to be addressed in a quantitative way and offer nice perspective for future works. 

\section*{Acknowledgment}

My warm thanks to Andr\'e Maeder whose enlightened theoretical developments allowed to explore
the effects of rotation in stellar models.

\end{document}